\documentclass[epj]{svjour}

\usepackage{graphicx}
\usepackage{fancyhdr}
\usepackage{amsmath}
\usepackage{cite}
\usepackage{graphicx}
\usepackage{amsfonts}
\usepackage{amssymb}%
\usepackage{epsf}
\usepackage{times}
\usepackage{a4wide}
\usepackage{epsfig}
\newcommand{\gev}{\,\mbox{GeV}}
\def\makeheadbox{{
 }}
\def\mytitle{The MSSM Higgs sector and $B-\overline{B}$ mixing for large $\tan\beta$} 
\def\myauthors{St\'ephanie Trine}    
\def\mytype{Contributed Talk}    
\def\mysession{Flavor Physics}

\begin{document}
\title{\boldmath The MSSM Higgs sector and $B-\overline{B}$ mixing for large $\tan\!\beta$\unboldmath}
\author{St\'ephanie Trine
\thanks{\emph{Email:} trine@particle.uni-karlsruhe.de}}                     
\institute{Institut f\"ur Theoretische Teilchenphysik,
        Karlsruhe Institute of Technology,
        D-76128 Karlsruhe, Germany}
\date{}
\abstract{
A systematic analysis of Higgs-mediated contributions to the $B_d$ and $B_s$ mass differences
is presented in the MSSM with large values of $\tan\beta$.
In particular, supersymmetric corrections to Higgs self-interactions are seen to modify the correlation between
$\Delta M_q$ and $\mathcal{B}(B_{q}\rightarrow\mu^{+}\mu^{-})$ for light Higgses.
The present experimental upper bound on $\mathcal{B}(B_{s}\rightarrow\mu^{+}\mu^{-})$ is nevertheless still sufficient to exclude
noticeable Higgs-mediated effects on the mass differences in most of the parameter space.
} 
\maketitle

\section{Introduction}
\label{intro}
\begin{figure}[t]
\begin{center}
\includegraphics[scale=.65,angle=0]{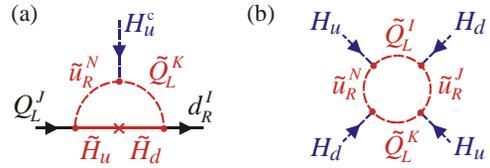}
\caption{(a) Squark-higgsino loop inducing a $\overline{d}_{R}^{I}\,Q_{L}^{J}\!\cdot\! H_u^c$ effective coupling.
The resulting quark mass matrix and neutral Higgs-quark vertices are not diagonalized simultaneously anymore, and FCNC are generated.
(b) Squark loop inducing an effective $U(1)_{PQ}$-violating Higgs self-coupling $(H_{u}\!\cdot\! H_{d})^{2}$.}
\label{fig:Matching} 
\end{center}
\end{figure}

If supersymmetry (SUSY) were exact, the two Higgs doublets of the Minimal
Supersymmetric extension of the Standard Model (MSSM) would not be able to mix,
and one of them only, $H_{u}=(h_u^+,h_u^0)$, would couple to up-type quark singlets
while the other one, $H_{d}=(h_d^{0*},-h_d^-)$, would interact with down-type quark singlets.
As SUSY-breaking is required to be soft, this peculiar Yukawa structure actually
holds at tree-level, and the dangerous flavour-changing neutral currents
(FCNC) that can be generated after spontaneous electroweak symmetry breaking by the coupling
of the quarks to the ``wrong'' Higgs are loop-suppressed:%
\begin{eqnarray}   \label{eq:1}
\mathcal{L}_{FCNC}^{Higgs}
&=&\kappa_{IJ}\ \overline{d}_{R}^Id_{L}^J
\left(c_\beta h_{u}^{0\ast}-s_\beta h_{d}^{0\ast}\right)
\nonumber\\
&+& \kappa_{JI}^{\ast}\ \overline{d}_{L}^Id_{R}^J
\left(c_\beta h_{u}^{0}-s_\beta h_{d}^{0}\right)  
\end{eqnarray}
in the quark mass eigenstate basis, with the abbreviations $c_\beta\equiv\cos\beta$, $s_\beta\equiv\sin\beta$,
and, under the Minimal Flavour Violation (MFV) assumption,%
\begin{eqnarray}   \label{eq:2}
\kappa_{IJ}\sim \frac{m_{I}}{v}\,V_{tI}^{\ast}V_{tJ}\ t_\beta^{2}\ \varepsilon_{Y},%
\end{eqnarray}
with $\varepsilon_{Y}$, a loop factor,
$t_\beta\equiv v_{u}/v_{d}$, the ratio of the two Higgs vacuum expectation values (VEV), and $v^2\equiv v_u^2+v_d^2$.
Note that the local effective interaction Eq.(\ref{eq:1}) supposes the scale hierarchy $M_{SUSY}\!\!\gg\!v$.
The loop factor $\varepsilon_{Y}$ is then essentially driven by squark and higgsino intermediate states (see Fig.\ref{fig:Matching}a).
Its effect is however non-decoupling in the limit $M_{SUSY}\rightarrow\infty$ as the induced effective operator has dimension-four
(see e.g. Refs.\cite{BabuK99,BurasCRS01,IsidoriR01,BurasCRS02} for details).
For large $t_\beta$, one can see that the loop suppression is compensated, opening the door to large Higgs-mediated
effects in flavour physics  \cite{BabuK99,Isidori07}.

\begin{figure}[t]
\begin{center}
\includegraphics[scale=0.45,angle=0]{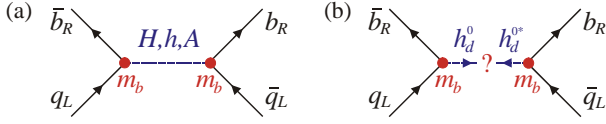}
\caption{(a) A priori dominant Higgs-mediated contribution to $\Delta M_{q}$.
(b) This contribution violates the $U(1)_{PQ}$ symmetry of the leading-order effective 2HDM for $v_d\to 0$.}
\label{fig:DeltaMCancellation} 
\end{center}
\end{figure}

A clean signature of this scenario was proposed in Ref.\cite{BurasCRS01}, which predicted a decrease
of the mass difference in the $B_{s}-\overline{B}_{s}$ system, $\Delta M_{s}$, with respect to
its Standard Model value, in direct correlation with an increase
of the $B_{s}\rightarrow\mu^{+}\mu^{-}$ branching fraction.
Interestingly, as first noted
in  \cite{Hamzaoui98,BabuK99}, the a priori dominant Higgs-mediated contribution to $\Delta M_{q}$
(see Fig.\ref{fig:DeltaMCancellation}a, $q=d,s$),%
\begin{eqnarray}   \label{eq:3}
\Delta M_{q}^{RR}\sim -\kappa_{bq}^2\left(  \frac{s_{\alpha-\beta}^{2}}{M_{H}^{2}}
+\frac{c_{\alpha-\beta}^{2}}{M_{h}^{2}}-\frac{1}{M_{A}^{2}}\right),
\end{eqnarray}
where $\alpha$ denotes the CP-even Higgs mixing angle and\linebreak $M_{H,h,A}$, the neutral Higgs masses,
actually vanishes when tree-level Higgs mass relations are implemented.
The aforementioned correlation was then derived flipping the chirality of one of the external $b$
quarks:%
\begin{eqnarray}   \label{eq:4}
\Delta M_{q}^{LR}\sim -\kappa_{bq}\kappa_{qb}^{\ast}\left(  \frac{s_{\alpha-\beta}^{2}}{M_{H}^{2}}
+\frac{c_{\alpha-\beta}^{2}}{M_{h}^{2}}+\frac{1}{M_{A}^{2}}\right),
\end{eqnarray}
which costs a factor of $\kappa_{qb}^{\ast}/\kappa_{bq}=m_{q}/m_{b}$.
The subject of the work reported here \cite{GorbahnJNT07} is the
systematic identification and computation of all contributions that
present one suppression factor with respect to the superficially dominant
term Eq.(\ref{eq:3}), and should thus be added to Eq.(\ref{eq:4}) before concluding on the correlation between
$\Delta M_{q}$ and $B_{q}\rightarrow\mu^{+}\mu^{-}$.

\section{\boldmath$\Delta M_q$ anatomy for large $\tan\!\beta$\unboldmath}
\label{sec:1}
In order to properly identify the relevant contributions, let us have a closer look at the cancellation in Eq.(\ref{eq:3}).
The tree-level Higgs mass matrices follow from the potential
\begin{eqnarray}   \label{eq:5}
&V^{(0)}=m_1^2H_{d}^{\dagger}H_{d}+m_2^2H_{u}^{\dagger}H_{u}+B\mu\left\{H_{u}\!\cdot\! H_{d}+h.c.\right\}\hspace{2mm}
\nonumber\\
&+\,\frac{{g}^{2}+g^{\prime 2}}{8}(H_{d}^{\dagger}H_{d}-H_{u}^{\dagger}H_{u})^{2}
+\frac{g^{2}}{2}(H_{u}^{\dagger}H_{d})(H_{d}^{\dagger}H_{u}),\hspace{7mm}
\end{eqnarray}
where $m_{1(2)}^2\equiv\left\vert\mu\right\vert^{2}+m_{H_{d(u)}}^{2}$,
$B\mu$ and $m^2_{H_{u,d}}$ denote soft-breaking terms, $\mu$ is the supersymmetric Higgs mass parameter,
and $H_{u}\!\cdot\! H_{d}\equiv H_u^T \varepsilon H_d$ with $\varepsilon^{12}=+1$.
In particular, we have:
\begin{eqnarray}
M_A^2=B\mu\ s_\beta^{-1}c_\beta^{-1}.
\end{eqnarray}
Consequently, for large $t_\beta$ (that is to say, $v_d\to 0$) and fixed $M_A$, $B\mu$
tends to zero, and, as the $h_u^0$ FCNC coupling in Eq.(\ref{eq:1}) also vanishes,
the two-Higgs doublet model (2HDM) composed of Eqs.(\ref{eq:1}) and (\ref{eq:5})
becomes invariant under the Peccei-Quinn-type symmetry with charge assignments \cite{IsidoriR01,HallRS93,DAmbrosioGIS02}
\footnote{Note that this symmetry is not spontaneously broken for $v_d\to 0$.}:
\begin{eqnarray}   \label{eq:6}
U(1)_{PQ}:\ \ Q(H_d)=Q(d_R^I)=1,\ \ Q(\mathrm{other})=0.
\end{eqnarray}
The cancellation in Eq.(\ref{eq:3}) now follows (at least in the $v_d\to 0$ limit) from the fact that the corresponding amplitude,
with two right-handed external $b$ quarks, requires a change of the PQ charge by two units (see Fig.\ref{fig:DeltaMCancellation}b),
and therefore cannot be generated by tree-level Higgs exchanges.

\begin{figure}[t]
\begin{center}
\includegraphics[scale=0.45,angle=0]{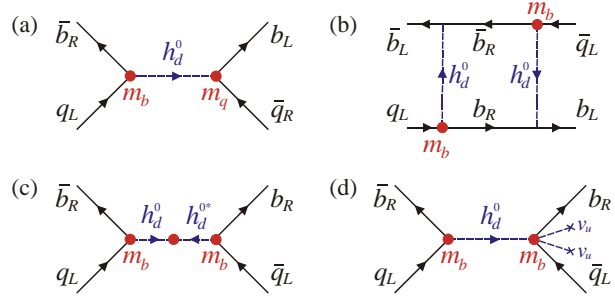}
\caption{Higgs-mediated contributions to $\Delta M_q$, resulting from (a) a chirality flip; (b) a weak scale loop;
(c) a SUSY loop in the Higgs potential; (d) Higher-dimension operators.}
\label{fig:DeltaMContributions} 
\end{center}
\end{figure}

Non-zero contributions are then obtained
\begin{itemize}
\item
either allowing the conservation of the PQ charge, which can be done by
(i) flipping the chirality of one of the external $b$ quarks, as said before (Fig.\ref{fig:DeltaMContributions}a) \cite{BurasCRS01};
(ii) avoiding the suppressed $\overline{b}_{L}q_{R}h^{0}_d$ FCNC coupling but allowing for one loop in the effective 2HDM
(Fig.\ref{fig:DeltaMContributions}b).
The diagram corresponding to this second possibility is readily computed from Eqs.(\ref{eq:1}) and (\ref{eq:5})
in the large $t_\beta$ limit, and found numerically small.
Note that charged Higgs effects are suppressed under our approximations \cite{BurasCRS01,IsidoriR01}.
\smallskip\item
or providing a breaking of the PQ symmetry via
(iii) sparticle-loop corrections to the tree-level effective potential $V^{(0)}$ (Fig.\ref{fig:DeltaMContributions}c);
(iv) higher-dimension quark-\linebreak Higgs effective operators (Fig.\ref{fig:DeltaMContributions}d).
These cost a SUSY loop, like the dimension-four effective coupling of\linebreak Fig.\ref{fig:Matching}a.
Then, as the only place where this loop can be compensated by a large $t_\beta$ factor is the modification
of the expression of the quark interaction eigenstates in terms of the quark mass eigenstates
in the tree-level $\overline{d}_{R}^I\,Q_{L}^J\!\cdot\! H_d$ Yukawa vertex
(here the quark fields are understood in the interaction eigenstate basis),
we actually end up again, in good approximation, with a coupling of the type $\overline{b}_{R}q_{L}h^{0\ast}_d$
(in the mass eigenstate basis now), and the usual cancellation mechanism takes place.
\end{itemize}
Note that higher-order sparticle-loop effects in the four-dimensional Yukawa vertices would still lead to a vanishing $\Delta M_q^{RR}$
at all orders in the $t_\beta^{-1}$ expansion for tree-level Higgs exchanges
as the combination of Higgs fields appearing in Eq.(\ref{eq:1}) would be unchanged.
Indeed, the occurence of $h_{u,d}^0$ or $h_{u,d}^{0*}$ is fixed by gauge symmetry for general dimension-four
Yukawa interactions, and, replacing $h_{u,d}^0$ by $v_{u,d}$, one must obtain zero in the quark mass eigenstate basis.

Corrections (iii) have been analyzed recently \cite{FreitasGH07,Parry06}.
Their size is however subject to controversy.
We thus go through them again in the next section.

\section{SUSY~corrections~to~the~Higgs~potential}
\label{sec:2} 
Sparticle loop corrections to $V^{(0)}$ are determined at the one-loop level
via a matching calculation on the most general dimension-four 2HDM potential for $M_{SUSY}\!\!\gg\!v$:
\begin{eqnarray}   \label{eq:7}
&V^{(1)}=m_{11}^{2}H_{d}^{\dagger}H_{d}
+m_{22}^{2}H_{u}^{\dagger}H_{u}
+\left\{  m_{12}^{2}H_{u}\!\cdot\! H_{d}+h.c.\right\}
\nonumber\\
&+\,\frac{\lambda_{1}}{2}(H_{d}^{\dagger}H_{d})^{2}
+\frac{\lambda_{2}}{2}(H_{u}^{\dagger}H_{u})^{2}
+\lambda_{3}(H_{u}^{\dagger}H_{u})(H_{d}^{\dagger}H_{d})
\nonumber\\
&+\,\lambda_{4}(H_{u}^{\dagger}H_{d})(H_{d}^{\dagger}H_{u})%
+\left\{ \frac{\lambda_{5}}{2}(H_{u}\!\cdot\! H_{d})^{2}\right.   \hspace{18mm}
\nonumber\\
&-\,[\lambda_{6}(H_{d}^{\dagger}H_{d})+
\lambda_{7}(H_{u}^{\dagger}H_{u})]\left. \strut (H_{u}\!\cdot\! H_{d}) 
+h.c.\right\},   \hspace{7mm}\label{2.1}
\end{eqnarray}
where $m_{12}^{2}$ and $\lambda_{5,6,7}$ may be complex.
Such a computation was actually already performed in the context of the corrections to the lightest Higgs mass $M_h$.
The explicit expressions for the $\lambda$'s available in the literature \cite{Lambdas}, however, assume
various approximations such as degenerate squark soft-breaking parameters
or real trilinear terms. These were removed in the computation of the Higgs mass matrices \cite{Mh}, but an updated list of
$\lambda$'s including the effects of all sparticles for arbitrary flavour structure has to our knowledge not been published.
Yet $\Delta M^{RR}_q$ in Eq.(\ref{eq:3}) takes a particularly transparent form when working in the Higgs interaction eigenstate basis,
being directly related to the $U(1)_{PQ}$-violating Higgs self-couplings $\lambda_{5}$ and $\lambda_{7}$, see Eq.(\ref{eq:12}).
We thus performed the matching again, keeping arbitrary $3\!\times\! 3$ soft-breaking matrices.
Particular attention was paid to the definition of the Higgs fields in the effective 2HDM, closely related to the definition of $t_\beta$, as we now briefly explain.

We chose to renormalize the MSSM parameters $m^2_{1}$ and $m^2_2$ such that
the one-loop tadpoles are renormalized to zero at the matching scale, i.e.,
the tree-level VEV $v^{(0)}_{u,d}$ still sit at the minimum
of the potential (no finite counterterms are introduced for the parameters $m_{ij}^2$ and $\lambda_i$ in the effective 2HDM).
The actual one-loop VEV $v_{u,d}^{(1)}$ must however take into account the field redefinition
needed to cast the kinetic terms
\begin{eqnarray} \label{eq:8}
\mathcal{L}_{Kin}^{(1)}&=&(1+\delta Z_{dd})\partial_{\mu}H_{d}^{\dagger}\partial^{\mu}H_{d}
\nonumber\\
&+&(1+\delta Z_{uu})\partial_{\mu}H_{u}^{\dagger}\partial^{\mu}H_{u}
\nonumber\\
&-&\left\{ \strut \delta Z_{ud}\,\partial_{\mu}H_{u}\!\cdot\!\partial^{\mu}H_{d}+h.c.\right\}
\end{eqnarray}
induced by the matching of the two-point Green functions into the canonical form.
We then have:
\begin{eqnarray} \label{eq:9}
&\left(\begin{array}[c]{c}%
v_{u}^{(1)}\\
v_{d}^{(1)}%
\end{array}\right)
=(1_{2\times 2}+\frac{\delta Z}{2})
\cdot
\left(\begin{array}[c]{c}%
v_{u}^{(0)}\\
v_{d}^{(0)}
\end{array}\right),
\end{eqnarray}
where $\delta Z_{21}\equiv\delta Z_{ud},\ \delta Z_{12}\equiv\delta Z_{ud}^\ast$, etc. Now, we take advantage of the freedom to change the Higgs basis \cite{HiggsBasis},
\begin{eqnarray} \label{eq:10}
\left(\begin{array}[c]{c}%
H_{u}^{\prime}\\
-H_{d}^{c^{\prime}}%
\end{array}\right)  
=e^{i\delta H/2}
\cdot
\left(\begin{array}[c]{c}%
H_{u}\\
-H_{d}^{c}
\end{array}\right),
\end{eqnarray}
where $\delta H$ is an arbitrary $2\!\times\! 2$ hermitian matrix and $H_d^c\equiv \varepsilon H_d^{*}$, to
(i) keep the VEV real and positive
(ii) more importantly, ensure that the corrections to $t_\beta$ are $t_\beta$-suppressed,
or, in other words, that $v_d$ does not receive any corrections from $v_u$.
This amounts to the following modification of Higgs field redefinition:
\begin{eqnarray} \label{eq:11}
&\frac{\delta Z}{2}\to\frac{\delta Z+i\delta H}{2}\!=\!
\left(\begin{array}[c]{cc}
\frac{\delta Z_{uu}}{2}+i t_\beta^{-1}\text{Im}(\delta Z_{ud}) & \,\delta Z_{ud}^{\ast}
\\0 &\frac{\delta Z_{dd}}{2}
\end{array}\right). \hspace{5mm}
\end{eqnarray}

The effects of the corrected Higgs masses and mixings on the ``flipped'' amplitude Eq.(\ref{eq:4})
are not essential, and we will ignore them here for simplicity. In the large $t_\beta$ limit, we then have:
\begin{eqnarray}   \label{eq:12}
&\Delta M_{q}^{LR}\sim -\kappa_{bq}\kappa_{qb}^{\ast} \frac{2}{M_A^2}.
\end{eqnarray}
In contrast, these effects are of course crucial for the ``non-flipped'' amplitude Eq.(\ref{eq:3}),
given for large $t_\beta$ to a good approximation by:
\begin{eqnarray}   \label{eq:13}
&\Delta M_{q}^{RR}\sim \kappa_{bq}^2 \left( \lambda_{5}-\lambda_{7}^{2}/\lambda_{2}\right) \frac{v^2}{M_A^4}
\end{eqnarray}
in the absence of new CP-violating phases. The above quantity is generated via the PQ-symmetry breaking
brought about by the $\mu$ parameter at loop-level.
To be explicit, in the case of $\lambda_5$, we obtain (within MFV and discarding the small contributions from the first two generations,
as well as those proportional to $g^\prime$):
\begin{eqnarray} \label{eq:14}
\lambda_{5}=
-\frac{3y_{t}^{4}}{8\pi^{2}}\frac{a_{t}^{2}\mu^{2}}{M_{\widetilde{t}R}^{4}}
L_{1}\biggl(\frac{M_{\widetilde{t}L}^{2}}{M_{\widetilde{t}R}^{2}}\biggr)
-\frac{3y_{b}^{4}}{8\pi^{2}}\frac{a_{b}^{2}\mu^{2}}{M_{\widetilde{b}R}^{4}}
L_{1}\biggl(\frac{M_{\widetilde{t}L}^{2}}{M_{\widetilde{b}R}^{2}}\biggr)\hspace{1mm}
\nonumber\\
-\frac{y_{\tau}^{4}}{8\pi^{2}}\frac{a_{\tau}^{2}\mu^{2}}{M_{\widetilde{\tau}R}^{4}}
L_{1}\biggl(\frac{M_{\widetilde{\tau}L}^{2}}{M_{\widetilde{\tau}R}^{2}}\biggr)
+\frac{3g^{4}}{8\pi^{2}}\frac{\mu^{2}}{M_{\widetilde{W}}^{2}}
L_{1}\biggl(\frac{\mu^{2}}{M_{\widetilde{W}}^{2}}\biggr),\hspace{7mm}\label{3.24}
\end{eqnarray}
with the loop function
\begin{eqnarray} \label{eq:15}
L_{1}(x)=\frac{-1}{(1-x)^{2}}-\frac{(1+x)\ln x}{2(1-x)^{3}}.\label{3.26}%
\end{eqnarray}
A typical contribution is depicted in Fig.\ref{fig:Matching}b.

In Ref.\cite{Parry06}, the corrected masses and mixings in Eq.(\ref{eq:3}) were determined
using the FeynHiggs package. We disagree numerically with the obtained results.
We also do not reproduce the pole singularity for $M_h\!\!=\!\!M_H$ found in Ref.\cite{FreitasGH07}.
From our analysis, it emerges that the source of the non-vanishing of $\Delta M^{RR}_q$ 
is to be found in the Higgs self-couplings $\lambda_5$ and $\lambda_7$ for large $t_\beta$,
related to CP-even Higgs mixing self-energies.

\section{Numerical analysis}
\label{sec:3} 
\begin{figure*}[t]
\begin{center}
\includegraphics[scale=0.97,angle=0]{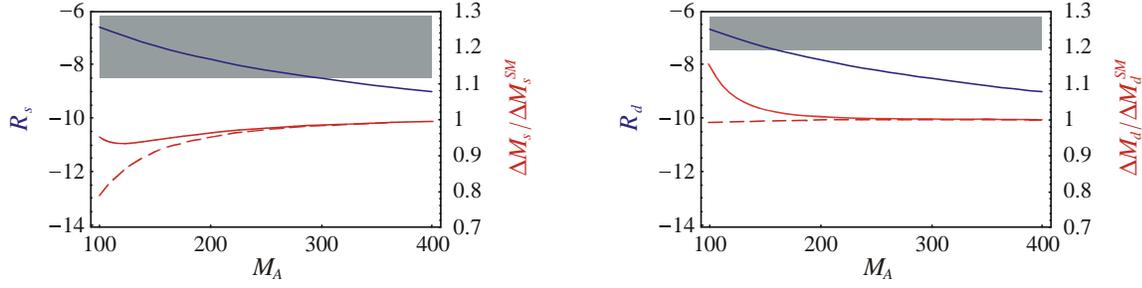}
\caption{Left: constraint from $\mathcal{B}(B_{s}\rightarrow\mu^{+}\mu^{-})$ on $\Delta M_s$.
The dark gray (blue) line is the theoretical prediction for $R_s\equiv\log_{10}[\mathcal{B}(B_{s}\rightarrow\mu^{+}\mu^{-})/\Delta M_s]$,
the light gray (red) lines indicate the size of SUSY effects in $\Delta M_s$, and the gray band shows the values of $R_s$
excluded experimentally \cite{Exps}.
The dashed line corresponds to $\Delta M_s=\Delta M_s^{SM}+\Delta M_s^{LR}$, while the plain line also takes $\Delta M_s^{RR}$ into account.
Supersymmetric parameters have been fixed as follows:
$t_\beta\!=\!40,\ a_{t,b}\!=\!2000\gev,\ M_{\widetilde{g}}\!=\!\mu\!=\!1500\gev,\ M_{\widetilde{q}}\!=\!M_{\widetilde{W}}\!=\!1000\gev,\ M_{1}\!=\!500\gev$.
Right: Analogue for the correlation between $\Delta M_d$ and $\mathcal{B}(B_{d}\rightarrow\mu^{+}\mu^{-})$
(experimental values from \cite{Exps,Expd}).
The bound on $\mathcal{B}(B_{d}\rightarrow\mu^{+}\mu^{-})$ is at present not as efficient as the bound on $\mathcal{B}(B_{s}\rightarrow\mu^{+}\mu^{-})$ to exclude Higgs-mediated effects on the mass differences, and $R_s$ ($\simeq R_d$) is preferably used.
}
\label{fig:DeltaMMA} 
\end{center}
\end{figure*}

As we already mentioned, Eq.(\ref{eq:12}) is responsible for a decrease of the $B_{s}-\overline{B}_{s}$ mass difference,
while $\Delta M_d$ is basically unaffected due to $\kappa_{db}\sim m_d$ \cite{BurasCRS01}:
\begin{eqnarray} \label{eq:16}
\Delta M_{q}^{LR}=C_{q}^{LR}X
\left[\frac{m_{s}}{0.06\gev}\right]
\left[\frac{m_{b}}{3\gev}\right]
\left[\frac{P_{2}^{LR}}{2.56}\right] \hspace{4mm}
\end{eqnarray}
with $C_{s}^{LR}=-14$ ps$^{-1}$, $C_{d}^{LR}\sim0$ ps$^{-1}$, and
\begin{eqnarray} \label{eq:17}
X=\frac{\left(  \varepsilon_{Y}16\pi^{2}\right)  ^{2}}
{\left(  1+\widetilde{\varepsilon}_{3}t_\beta\right)  ^{2} \left(  1+\varepsilon_{0} t_\beta\right)  ^{2}}
\frac{m_{t}^{4}}{M_{W}^{2}M_{A}^{2}}\left[  \frac{t_\beta}{50}\right]  ^{4}. \hspace{1mm}
\end{eqnarray}
The loop factors $\varepsilon_0$, $\varepsilon_Y$ and $\widetilde{\varepsilon}_{3}\equiv\varepsilon_0+y_t^2\varepsilon_Y$
may be found in Refs.\cite{BurasCRS02,FreitasGH07}, including the effects of the electroweak couplings $g$ and $g^\prime$.
The new contribution Eq.(\ref{eq:13}), on the other hand, increases both $\Delta M_s$ and $\Delta M_d$
(note that $\lambda_5$ and the bag factor $P_1^{SLL}$ are both negative):
\begin{eqnarray} \label{eq:18}
\Delta M_{q}^{RR}=C_{q}^{RR}X
\left[\frac{m_{b}}{3\gev}\right]^{2}
\left[\frac{P_{1}^{SLL}}{-1.06}\right] \hspace{5mm}
\nonumber\\
\times\frac{M_{W}^{2}}{M_{A}^{2}}\left(-\lambda_{5}+\lambda_{7}^{2}/\lambda_{2}\right)  16\pi^{2} \hspace{11mm}
\end{eqnarray}
with $C_{s}^{RR}=+4.4$ ps$^{-1}$ and $C_{d}^{RR}=+0.13$ ps$^{-1}$.
The numbers in Eqs.(\ref{eq:16}) and (\ref{eq:18}) have been obtained using
$|V_{ts}V_{tb}^{\ast}|=0.041$ \cite{BallF06}, $|V_{td}V_{tb}^{\ast}|=0.0086$ \cite{BallF06},
$F_{B_s}=0.24\gev$ and $F_{B_d}=0.2\gev$. These values suffer from large uncertainties, and are given here for the purpose of illustration
(ratios are defined for actual numerical studies, see Fig.\ref{fig:DeltaMMA}).
They correspond to the Standard Model central values
$\Delta M_s^{SM}=20$ ps$^{-1}$ and $\Delta M_d^{SM}=0.59$ ps$^{-1}$.

A first observation is that the typical effect of $\Delta M_s^{RR}$ is suppressed with respect to that of $\Delta M_s^{LR}$,
which is due to a 1/2 symmetry factor and the small value of $P_1^{SLL}$.
The effective couplings in Eq.(\ref{eq:18}) are also not very large. To get an idea of their size,
the residual $\lambda_5$ value for $M_{SUSY}\to\infty$ is given by
\begin{eqnarray}
&\lambda_5\to-\frac{1}{2}(y_t^4+y_b^4+\frac{y_\tau^4}{3}-g^4) \frac{1}{16\pi^2}.
\end{eqnarray}
The ``non-flipped'' contribution $\Delta M_q^{RR}$ can still be relevant for small $M_A$ (i.e., $<200\gev)$.
However, in that case, the experimental upper bound on $\mathcal{B}(B_{s}\rightarrow\mu^{+}\mu^{-})$ \cite{BabuK99,BurasCRS01,IsidoriR01,BurasCRS02,BqMuMu}
imposes tough constraints on $X$ and $t_\beta$,
\begin{eqnarray} \label{eq:19}
\mathcal{B}(B_{q}\rightarrow\mu^{+}\mu^{-})=C_{q}X\frac{M_{W}^{2}}{M_{A}^{2}%
}\left[  \frac{t_\beta}{50}\right]  ^{2}
\end{eqnarray}
with $C_{s}=3.9\ 10^{-5}$ and $C_{d}=1.2\ 10^{-6}$, suppressing the overall effect in $\Delta M_q$ (see Fig.\ref{fig:DeltaMMA}).
In other words, the correlation between $\mathcal{B}(B_{q}\rightarrow\mu^{+}\mu^{-})$ and $\Delta M_q$ can be modified,
but this does not spoil the conclusion derived in Refs.\cite{Correlation} that the present data on $\mathcal{B}(B_{s}\rightarrow\mu^{+}\mu^{-})$
already exclude visible effects in $\Delta M_s$ (it actually reinforces it, see Fig.\ref{fig:DeltaMMA}),
while a similar conclusion can be reached for $\Delta M_d$.

Non negligible effects compatible with the $B_{q}\rightarrow\mu^{+}\mu^{-}$ constraints are not excluded
in some corners of parameter space, for large $\mu$ and large $a$-terms. However, they again require light Higgses,
which is in any case disfavored (and partly excluded) by the observed $B\to\tau\nu$ branching fraction.
A small window for very light CP-odd Higgs mass
is still allowed for large $t_\beta$, but corresponds to the somewhat fine-tuned scenario where charged Higgs effects
in $B\to\tau\nu$ interfere destructively with the Standard Model amplitude, and are about twice its value.

\section{Conclusion}
\label{ccl} 
We have performed a systematic analysis of Higgs-mediated contributions to $\Delta M_q$ in the MFV-MSSM with large $\tan\beta$ and sparticles at the TeV scale.
For $M_A>200\gev$, no new effect is found.
For small $M_A$, SUSY loop corrections to the Higgs self-interactions can (moderately) modify the correlation between $\Delta M_q$ and $\mathcal{B}(B_{q}\rightarrow\mu^{+}\mu^{-})$. The present experimental upper bound on $\mathcal{B}(B_{s}\rightarrow\mu^{+}\mu^{-})$ is however still sufficient to exclude
visible Higgs-mediated effects on $\Delta M_q$  in (practically) all parameter space.
The precise measurements of $\Delta M_q$ are then to be used more as a normalization to avoid the large uncertainties
related to $F_{B_q}$ and $V_{tq}$ when using $B_{q}\rightarrow\mu^{+}\mu^{-}$ to probe the MSSM in the large $t_\beta$ regime.
\\[10mm]
\emph{Acknowledgments}:
It's a pleasure to thank my collaborators Martin~Gorbahn, Sebastian~J\"ager and Ulrich~Nierste.
This work was supported by the DFG grant No.~NI~1105/1--1, by the DFG--SFB/TR9,
and by the EU Contract No.~MRTN-CT-2006-035482, \lq\lq FLAVIAnet''.

\end{document}